# Data-Driven Design of a New Organic Semiconductor *via* an Electronic Structure Chart

Daniel M. Packwood[1], Yu Kaneko[2], Daiji Ikeda[2], Mitsuru Ohno[2]

[1] Institute for Integrated Cell-Material Sciences (iCeMS), Kyoto University, Kyoto 606-8501, Japan

[2] Daicel Corporate Research Center, Innovation Park (iPark), Daicel Coorporation, Himeiji, Japan

**Abstract**

Data-driven methodologies for designing new materials are developing apace, yet advances for organic crystals have been infrequent. For organic crystals, the need to predict solid-state electronic properties from molecular structure alone is an exceedingly difficult task for typical, regression-based design strategies. In this paper, we present a new strategy for designing organic crystals which circumvents the need to regress solid-state physical properties directly. At the core of this strategy is an electronic structure chart, a two-dimensional projection of an organic crystal database in which each material is positioned according to its solid-state electronic properties. We illustrate this strategy by identifying a new molecule which is predicted to show a targeted band gap and better-than-average band curvatures in the crystalline state. This strategy is the first data-driven method which can design new molecules on the basis of genuine solid-state electronic properties, and has potential to accelerate breakthroughs in the field of organic electronics and beyond.

## 1. Introduction

Computational materials science, like most other fields of science, has embraced a data-centric culture. Over the last few years the generation of material databases *via* high-throughput density functional theory (DFT) calculations, followed by the application of machine learning (ML) to them, has become standard protocol [1, 2]. Much of the attraction of this protocol lies in its potential for assisting the design of new materials, on the basis of structure-property relationships obtained by ML. However, while this potential has been vividly demonstrated for certain classes of materials including single molecules [3, 4, 5], molecular assemblies [6], polymers [7, 8, 9], and various types of inorganic solids [10, 11], significant difficulties have been encountered for other cases.

Organic crystals are comprised of organic molecules held in-place by non-covalent interactions. Their electronic structures, which arise from coupling between molecular orbitals of neighboring molecules, are typically semiconducting in character and often support ambipolar charge transport, making organic crystals attractive for device applications such as light-emitting diodes and field-effect transistors [12, 13]. While the best-performing organic crystals known at present have charge carrier mobilities several orders of magnitude smaller than their inorganic counterparts [14], there are high hopes that more competitive organic crystals remain undiscovered. This hope is inspired by the enormous range of molecules which synthetic chemists can potentially create and crystallise [15, 16]. It is here, however, where hope turns to pessimism. The intuitive approach to designing new organic crystals – making a sequence of small additions and adjustments to a promising initial structure in an attempt to optimise properties – is notoriously ineffective. Minor changes to molecule structure usually induce dramatic changes in crystal packing and molecular orbital shapes, causing electronic properties to change unpredictably. Such unpredictable changes make it difficult to tune electronic properties on the basis of small molecular structure adjustments. Alternative approaches to designing organic crystals are therefore urgently needed.

The complex relationship between molecule structures and crystal-phase electronic structures makes organic crystals a challenging target for ML-based materials design. Most ML-based design strategies involve the creation of regression models which map structural descriptors to the physical property of interest. While accurate regression models for band gaps and other properties of organic crystals have been built on the basis of *crystal structure-derived* descriptors [17, 18], the effectiveness of such models for designing new materials is necessarily limited. For the case of organic crystals, an important point must be observed: the regression model must exclusively map *molecule structure-derived descriptors* to crystal-phase electronic properties. This point follows from the fact that, in order to be useful for designing new materials, the regression model should only map from variables which can be controlled during the material synthesis process. Molecular structures are the only part of an organic crystal which a synthetic chemist has direct control over. This point has been recognised by several recent authors, who have devised regression models based on molecular-structure-derived descriptors and used them to predict molecules for new organic crystals. However, instead of predicting genuine crystal-phase electronic properties, these models predict molecule- or dimer-level properties such as reorganisation energy [19, 20, 21] and transfer integrals [22] instead. In order to use such regression models to predict molecules for new organic crystal materials, it must be assumed that the molecules pack in a specific way in the crystalline state, or that molecule-level properties (such as reorganisation energy or HOMO-LUMO gaps) have the overwhelming influence on the crystal-state electronic property of interest (such as charge carrier mobility or band gap). Neither assumption is satisfactory considering the unpredictable influence of molecule structure on crystal structure, as well as the complicated and poorly understood connection between bulk electronic properties and local molecular properties [23, 24]. On the other hand, it may be difficult to do better than this; the regression of crystal-phase electronic properties onto molecular descriptors alone may be unreasonably difficult given the complicated relationship between these variables.

In this paper, we present an alternative data-driven strategy for designing new organic crystals with targeted crystal-phase electronic properties. Our strategy is based upon an *electronic structure chart* (ESC) a projection of an organic crystal database onto a two-dimensional space in which the organic crystal structures are judiciously arranged according to their band gaps and band curvatures. While the ESC is interesting in itself, its practical value lies in the fact that it simplifies the construction of models needed to design new organic crystals. Indeed, we can select a region of the ESC which contains many crystals with desirable electronic properties, and then construct a simple classification model which predicts on whether or not a candidate molecule will belong to this region when crystallised. Mathematically, the construction of this classification model amounts of regressing a binary variable, rather than a continuous physical property, onto a set of molecular descriptors. Binary variables, being less sensitive to changes in model input than continuous ones, are preferable when constructing models for crystal-phase properties on the basis of molecular descriptors. Having built this classification model, new molecules can then be designed by simply generating a list of candidate molecules and noting the ones for which the classifier returns a positive prediction. We demonstrate this methodology here and predict a new and synthetically realisable molecule which appears to have a small band gap and better-than-average hole transport properties in the crystal state.

This paper is organised as follows. In section 2 we describe our dataset and introduce electronic fingerprints, which are a band structure descriptor needed for constructing the ESC. The ESC is introduced in section 3 and characterised in detail. In section 4, we show how the ESC can be used to design molecules for new organic crystals, and the crystal-phase electronic properties of a molecule obtained by this approach is analysed in Section 5. Discussion and concluding remarks are left to Section 6.

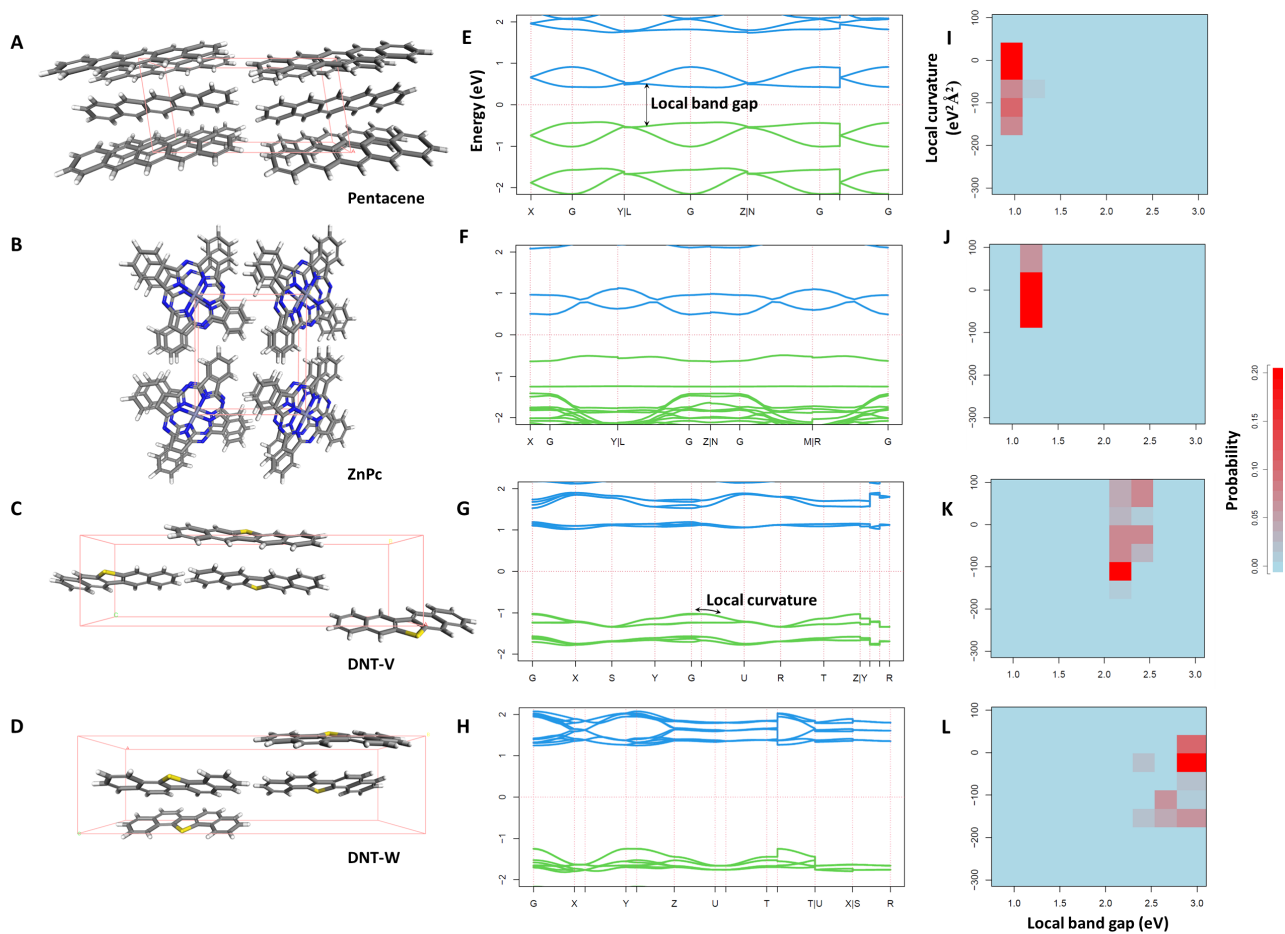

**Figure 1. Crystal structures, electronic band structures, and electronic fingerprints.** (A – D) Crystal structures of pentacene, ZnPc, DNT-V, and DNT-W, respectively. These are representative materials from our database. Grey, white, blue, purple, and yellow parts represent carbon, hydrogen, nitrogen, zinc, and sulfur atoms, respectively. Red lines indicate unit cell boxes. Images drawn with the Materials Studio Visualizer [62]. (E – H) Electronic band structures calculated for pentacene, ZnPc, DNT-V, and DNT-W, respectively. Occupied and unoccupied bands are coloured green and blue, respectively. The vertical thin red lines indicate Brillouin zone boundaries and the horizontal line indicates the Fermi level. Local band gap and local curvature are specific points along the Brillouin zone path are illustrated. (I – L) Electronic fingerprints for pentacene, ZnPc, DNT-V, and DNT-W, respectively. All plots were produced with R [51] using custom scripts.

## 2. Data collection and electronic fingerprints

A database of around 5500 crystal structures of organic crystals was compiled by random sampling from the Crystallography Open Database (COD) [25] (see "Database generation" in the Methods section). A full list of crystal structures used in this database is provided in the Supporting Information 1. Four representative structures from this database (pentacene, ZnPc, DNT-V [26], and DNT-W [27]) are shown in Figure 1 A – D. For each material in this database, electronic band structures were computed in a high-throughput manner *via* density functional theory (DFT). The band structures calculated for pentacene, ZnPc, DNT-V, and DNT-W are shown in Figure 1 E – H, respectively. These band structures, as with almost all materials in this database, show a clear band gap, indicating semiconductivity. While these band gaps will underestimate the experimental ones by 1 – 2 eV due to the generalized gradient approximation used in the DFT calculations [28], the band structure trends should be reliable.

It is well established that charge transport in organic crystals cannot be modeled quantitatively in the band transport picture due to the presence of strong electron-phonon coupling and polaron formation [29, 30, 31, 32, 33, 34, 35]. However, one must not confuse the use of band structures for modeling charge transport with the use of band structures for *building descriptors* for electronic properties. If we are to design new organic cystals with strong electronic coupling between neighboring molecules, then band structures are useful because the curvature of the bands reflects this electronic coupling (as can be seen in the tight-binding Hamiltonian constructed in a molecular orbital basis, whose off-diagonal elements induce band curvature). These same electronic coupling parameters usually appear as coefficients in advanced treatments of charge transport in organic crystals as well (for example, see the various mobility expressions in [33]). Band gaps also remain physically meaningful in the presence of electron-phonon coupling, although they tend to narrow slightly in the presence of thermal fluctuations of the nuclei positions [36]. While certainly not a satisfactory means of modeling room-temperature charge transport, band structures contain a wealth of information relevant to the design of new materials.

In order to create the electronic structure chart (ESC), descriptors must be extracted from the band structures. For each band structure, descriptors were extracted in three steps. First, $n$ $k$-points from the Brillouin zone were selected at random. Secondly, for each of these $k$-points, the local gap $\varepsilon_k$ (the energy difference between the highest occupied band and the lowest unoccupied band at that $k$-point) and the local curvature $c_k$ (the second derivative of the energy of the occupied band at that $k$-point) were calculated. This step results $n$ pairs of points ($\varepsilon_k, c_k$). In the third step, a joint probability distribution $P(\varepsilon, c)$ of local gaps and local curvatures across the Brillouin zone was created using the $n$ pairs of points obtained in the previous step (see Methods section "Electronic fingerprints and electronic structure chart generation" for further details). Examples of joint probability distributions obtained from the band structures of pentacene, ZnPc, DNT-V, and DNT-W are shown in Figure 1 I – L, respectively. These joint probability distributions are referred to as *electronic fingerprints*, and serve as descriptors for the electronic band structure. For a given organic crystal, the electronic fingerprint describes how the local gap and the local curvature jointly vary across the Brillouin zone. It therefore summarizes information on hole transport in the semiconductor as well as its optical properties at absolute zero temperature. These electronic fingerprints could easily be extended to describe electron transport by computing the curvature of the lowest unoccupied band.

Electronic fingerprints are potentially demanding to calculate, as a high-quality band structure over the entire Brillouin zone should be pre-calculated in order to obtain a uniform and reliable sample of local band gaps and curvatures. In this paper (including in the examples in Figure 1), we actually sample $k$-points from along the Brillouin zone path used in the band structure calculation in order to minimise computational demands. This approximation should be acceptable because the Brillouin zone paths used here, which were generated with the AFLOW package [37, 38], cover all of the representative parts of the Brillouin zone

The examples in Figure 1 demonstrate that the connection between molecule structure, crystal structure, and electronic structure is not simple. For example, pentacene and ZnPc have very different molecular and crystal structures, however their electronic fingerprints are somewhat similar. On the other hand, DNT-V and DNT-W possess very similar molecule and crystal structures, yet their electronic fingerprints are concentrated in different locations. This fact is consistent with experience, namely that crystal and electronic structures are extremely sensitive to molecule structure.

Construction of the ESC requires a measure of dissimilarity between pairs of band structures. Because electronic fingerprints are probability distribution functions, we can quantify these dissimilarities using techniques from probability theory. In this work, we measure the dissimilarity between pairs of electronic fingerprints using the Jansen-Shannon (JS) divergence [39]. Consider two organic crystals and their respective electronic fingerprints $P(\varepsilon, c)$ and $P'(\varepsilon, c)$. The JS divergence between $P$ and $P'$ is defined as

$$d(P, P') = \frac{1}{2}k(P, M) + \frac{1}{2}k(P', M). \qquad (1)$$

where $M(\varepsilon, c) = (P(\varepsilon, c) + P'(\varepsilon, c))/2$ and $k$ is the Kullback-Leibler divergence, defined as

$$k(P, M) = \int P(\epsilon, c) \log_{10} \frac{P(\epsilon, c)}{M(\epsilon, c)} d\epsilon dc, \qquad (2)$$

and similarly for $k(P', M)$. See Methods section "Electronic fingerprints and electronic structure chart generation" for calculation details. In principle, the JS divergence ranges between 0 and $\log_{10}(2) = 0.301$. However, for the cases studied here, we find that the JS distance varies between about 0.25 and 0.301 for most cases. In Figure 1, the JS divergence between pentacene and ZnPc is 0.28, implying moderate overlap of electronic fingerprints, and the distance between DNT-V and DNT-W is 0.30, implying weaker overlap. Between ZnPc and DNT-V it is 0.301, implying negligible overlap between electronic fingerprints and hence completely different electronic structures.

### 3. Electronic structure chart for organic crystals

The electronic structure chart (ESC) is presented in Figure 2A. It was computed using the tSNE (t-distributed stochastic neighbor embedding) method (see Methods section "Electronic fingerprints and electronic structure chart generation" for details) [40]. The tSNE method generates a two-dimensional projection of the database in such a way that the distances between nearby points correspond to the JS divergence in (1). In this diagram, each point represents one organic crystal structure from our database. Crystals which are close together in the ESC should therefore have similar band structures. Indeed, it can see that pentacene and ZnPc are located close together, whereas DNT-V and DNT-W are located further away.

Close inspection of the ESC reveals that it is organized on two levels. On a global level, the compounds appear to be organized according to their local gaps. This is clear from Figure 2A, in which the points are colored according to the average local gap of the material (the local gap averaged over the Brillouin zone path). Following the trail of points starting from the left-hand side, the points gradually change from red (indicating large gaps) to bright blue (indicating very small gaps). These two extremes respectively correspond to insulating, σ-bond-dominated componds to metal-containing compounds with small gaps due to partially filled metal ion orbitals. Aromatic compounds and other π-electron heavy compounds populate the ESC between these two extremes (see Supporting Information 2 for examples).

Locally within in the ESC, the organic crystals appear to be arranged according to local curvatures. Figure 2B presents the same ESC, but with the points colored according to the logarithm of the average local curvature (the absolute value of the local curvature averaged over the Brillouin zone path). While it appears that cases with lower curvature (blue points) tend to concentrate on the edges of the trail, and that cases with higher curvature tend to concentrate in the middle of the trail (red points), it is difficult to draw unambiguous conclusions on the distribution of curvatures from Figure 2B alone. In order to determine whether local curvatures are indeed distributed in a systematic, non-random fashion across the ESC, we computed a series of autocorrelation functions (ACFs) as shown in Figure 2C. Roughly speaking, an ACF measures how quickly a quantity of interest changes as we move through the trail of points in the ESC (see Methods section "Autocorrelation functions" for details). The ACF corresponding to random noise (blue line), which was obtained by assigning independent random variates from a standard normal distribution to each crystal, represents a base case (blue curve). In this case, ACF quickly decays to zero. This ACF can be taken as a control case, where the quantity of interest is randomly distributed across the ESC. When the quantity of interest is the average local curvature, the ACF decays appreciably more slowly (green line). This unambiguously shows that local curvatures are distributed in a non-random, systematic way across the ESC. When the quantity of interest is the average energy gap, the ACF decays extremely slowly (red), which reflects the slow variation of the band gap over the entire of the ESC. The stark difference in ACF decay rates for the cases of average local gaps and average local curvatures reflects the two-level, global versus local, distribution of these properties across the ESC.

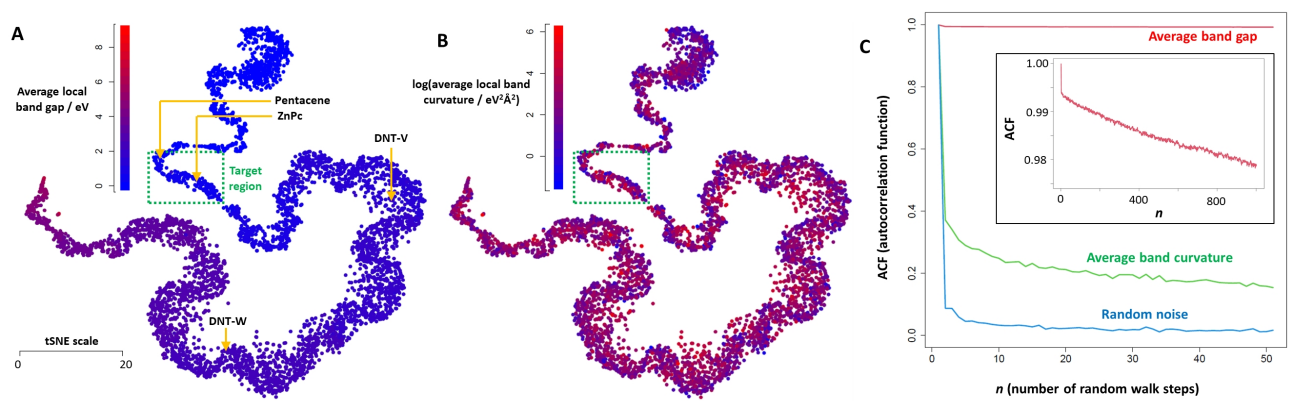

**Figure 2. Electronic structure chart**. (A) Electronic structure chart computed using the method described in the text. In the electronic structure chart, each material from the organic crystal database is projected into a 2D plane in such a way that crystals with similar electronic fingerprints are placed close together. Each point indicates one organic crystal from the database. The locations of pentacene, ZnPc, DNT-V, and DNT-W are indicated. The points are colored according to their average local band gap. (B) As for (A), but with points colored according to the logarithm of the average local valance band curvature. (C) Autocorrelation functions (ACFs) for the average band gap, average band curvature, and random noise. These were computed by simulating a random walk which hops between nearby points in the electronic structure chart. Insert shows the ACF for the average band gap over a longer range of random walk steps. Figures were produced using R with the Rtsne package [51, 52].

## 4. ESC-based material design strategy

Having shown that the ESC arranges the organic crystals in a systematic way, we now consider whether it can be used for designing materials. For this reason, we consider the region shown in the green box in Figure 2A (the 'target region'). This region contains organic semiconductors with small band gaps (around 1.0 – 2.0 eV), and also contains some quintessential, high hole mobility compounds (pentacene, ZnPc, and $C_{60}$). By harnessing the molecular structure information contained in the target region, we aim to design a new molecule which shows a small band gap and good band curvature in the crystalline state.

Figure 3 summarises our ESC-based material design procedure. It involves three steps, namely (i) building a molecule database, (ii) using this database to build a classification model which predicts whether or not a molecule will belongs to the target region upon crystallization, and (iii) generating a large list of molecules and predicting promising candidates using the regression models.

The molecule design problem is illustrated in Figure 3A. In short, we wish to predict a molecule which belongs to the target region (i.e., has "label 1") and has a high melting point so that it can operate in real device. The construction of the molecule database (step (i)) is described in Figure 3B. First, each crystal structure in our crystal database (described in section 2) was expanded into a 2 x 2 x 2 supercell, from which the isolated molecules were extracted. Cases in which multiple types of organic molecules, organic molecule radicals, or isolated inorganic ions (such as $K^+$) were ignored. This procedure yielded 2264 molecules in total. Each molecule was given a label 1 or 0, depending upon whether its parent crystal structure belonged to the target region or not. The molecule structures were stored as SMILES strings for subsequent processing (see Methods section "Prediction of candidate molecules" for details).

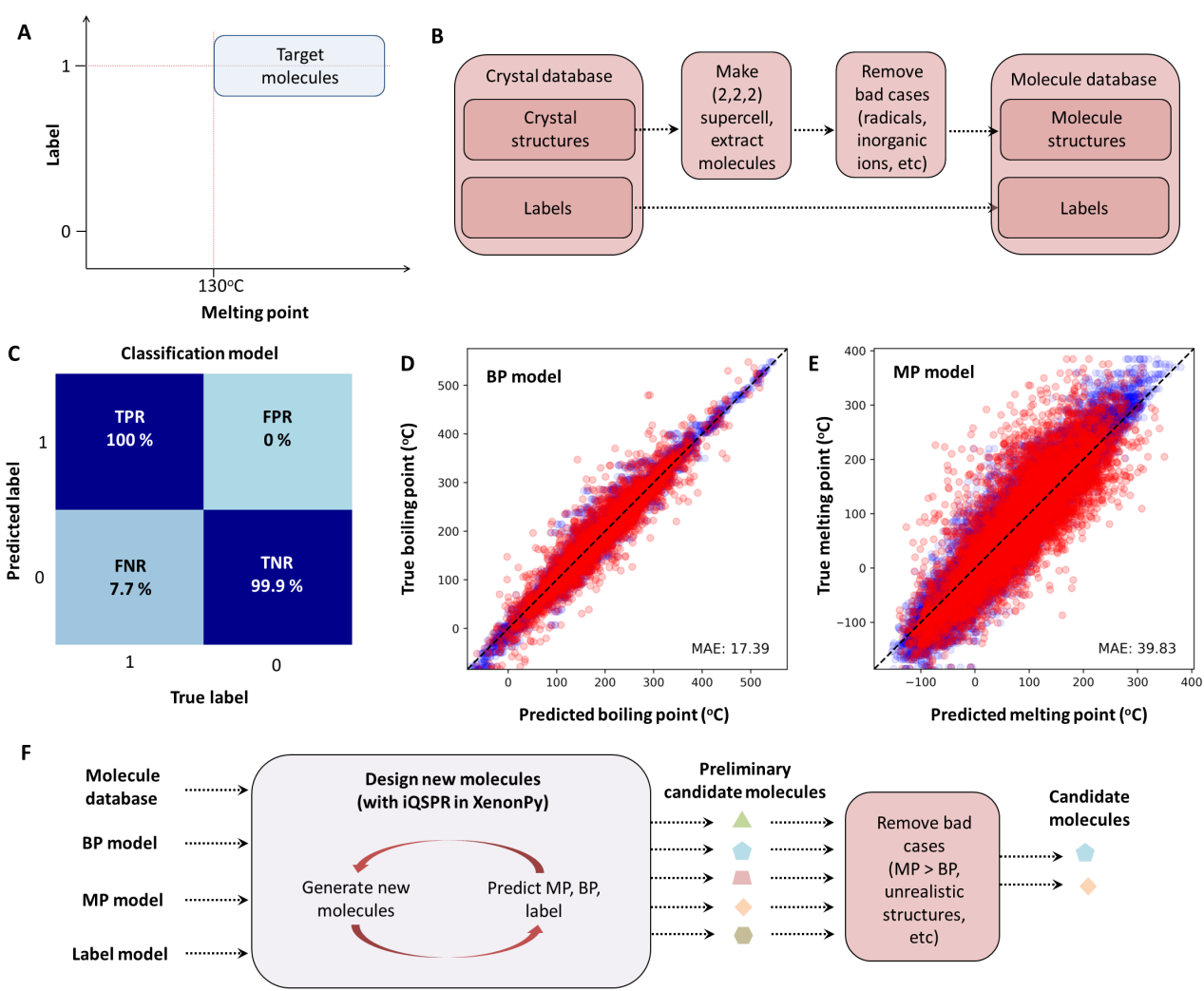

**Figure 3. Design of new molecules with targeted solid-state properties.** (A) Illustration of the molecule design space. The horizontal and vertical axes indicate solid-state properties, as predicted by continuous-valued regression models. Melting point is the predicted melting point in the crystalline state. Label indicates whether or not the molecule, in the crystalline state, resides within the target region of the electronic structure chart (green box in Figure 2). The goal of this part is to identify molecules within the blue box ("target molecules"). (B) Construction of the molecule database for constructing the regression models. The molecules are extracted from the organic crystal database in Figure 2. (C) Performance of the classification model, presented as a confusion matrix. TPR, FPR, FNR, and TNR are the true positive rate, false positive rate, false negative rate, and true negative rate, respectively. (D, E) Performance of regression models for predicting boiling point and melting point (E), respectively, for organic molecules the molecules in the crystalline state. In (C) and (D), blue and red points correspond to training and test data, respectively. (F) Process for predicting molecules with targeted solid-state properties. iQSPR within the XenonPy package [46] is used to predict candidate molecules on the basis of the molecule database and the models constructed in (C- E). Details are discussed in the text.

The performance of the classification model is shown in Figure 3C as a confusion matrix (step (ii)). This model was constructed using the molecule database from step 1, using ECFP descriptors with radius 4 [41] and the XGboost library [42]. This classification model was actually trained as regression model, with labels assigned based on whether or not the model output exceeded 0.5. The classification model achieved impressive accuracy, only misclassifying 7.7 % of the molecules in the test data. In addition to the classification model, we built two additional regression models for predicting the boiling points ('BP model') and melting points ('MP model'), respectively, of each molecule in the crystalline state (Figures 3D, 3E). We construct the BP and MP models in consideration of real device applications. An organic semiconducting material requires a melting point of at most 130°C in order to be compatible with device fabrication conditions. Moreover, by incorporating a regression model for the boiling point, we can avoid proposing unrealistic molecules in which the melting point exceeds the boiling point. The BP and MP models were trained using the data in reference [43], which provides SMILES strings and experimental boiling and melting points for thousands of organic compounds. The BP and MP models were constructed using pattern fingerprints (PatternFP) and extended connectivity fingerprints (ECFP) with radius 2, respectively, which were generated from the SMILES strings using the RDkit library [44]. Regression models were constructed using the XGboost library.

The prediction of candidate molecules (step (iii)) is performed using the Bayesian inverse molecular design algorithm (inverse-QSPR) in the XenonPy package [45, 46]. In short, this procedure takes the molecule database and above models as input, and generates a new list of candidate molecules based upon the ones in the database. The melting points, boiling points, and classification of the new molecules are then predicted with the above models, and the ones which have desirable properties (high melting points and classification 1) are noted. This procedure is iterated, each time using the results from the previous step to facilitate the generation of the candidate molecules. For details, see reference [46]. After 28 iterations we found that this procedure could consistently generate molecules with predicted label 1 (i.e., predicted to belong to the target region in the crystalline state). The final list contained 248 molecules (See Supporting Information 3).

## 5. Solid-state properties of the predicted organic semiconductor

While the molecules in the final list were chemically realistic, most of them contained a very large number of atoms and had very complex structures. Not only are these molecules difficult synthetic targets, but their predicted properties are unlikely to be reliable. This is because the machine learning models (Figure 3C - E) were constructed using training sets containing relatively simple structures, and are therefore not trained to make predictions on such complex structures. From the molecules in this final list, we therefore choose to focus on molecule **1** (Figure 4A insert) for further analysis. Molecule **1** has a relatively small, planar, and symmetric structure, similar to the many successful organic semiconducting compounds such as pentacene. Despite its simplicity, molecule **1** does not appear to have been reported elsewhere.

In order to rigorously evaluate molecule **1**'s physical properties in the crystalline state, it is necessary to know or predict its crystal structure. To the best of our knowledge, the crystal structure of molecule **1** is unknown. Moreover, crystal structure prediction for organic compounds is not yet routine, although impressive progress has been made over recent years [47]. Even if the crystal structure of molecule **1** could be easily predicted, there is no guarantee that the predicted structure would be the same as the one obtained in the laboratory, as particular polymorphs may by kinetically preferred under certain processing conditions. In order to estimate the electronic properties of molecule **1** in the crystalline state, we turn to a statistical procedure based upon principal component analysis (PCA) and an amorphous phase of molecule **1**. Here, PCA is used to generate descriptors based upon molecule- and dimer-level information. Crucially, these descriptors are shown to have moderate correlations with the band gaps and curvatures of the materials in the crystalline state. By extracting the molecule- and dimer-level information from an amorphous phase of molecule **1** and computing these descriptors, we can therefore estimate the electronic properties of molecule **1** in the crystalline state.

Figure 4A shows an amorphous phase of molecule **1** (referred to as amorphous **1**), which was generated using a quenched molecular dynamics simulation of 256 copies of molecule **1** in a large periodic cell (see Methods section "Generation of amorphous 1" for details). Since amorphous **1** resides in a local energy minimum, it is expected to exhibit many of the energy-minimising molecular packing modes which are present in the major crystal polymorphs of molecule **1**. We therefore extracted the 232 most energetically stable molecule pairs (dimers) from amorphous **1** for use in the subsequent steps (see Methods section "Dimer extraction" for details).

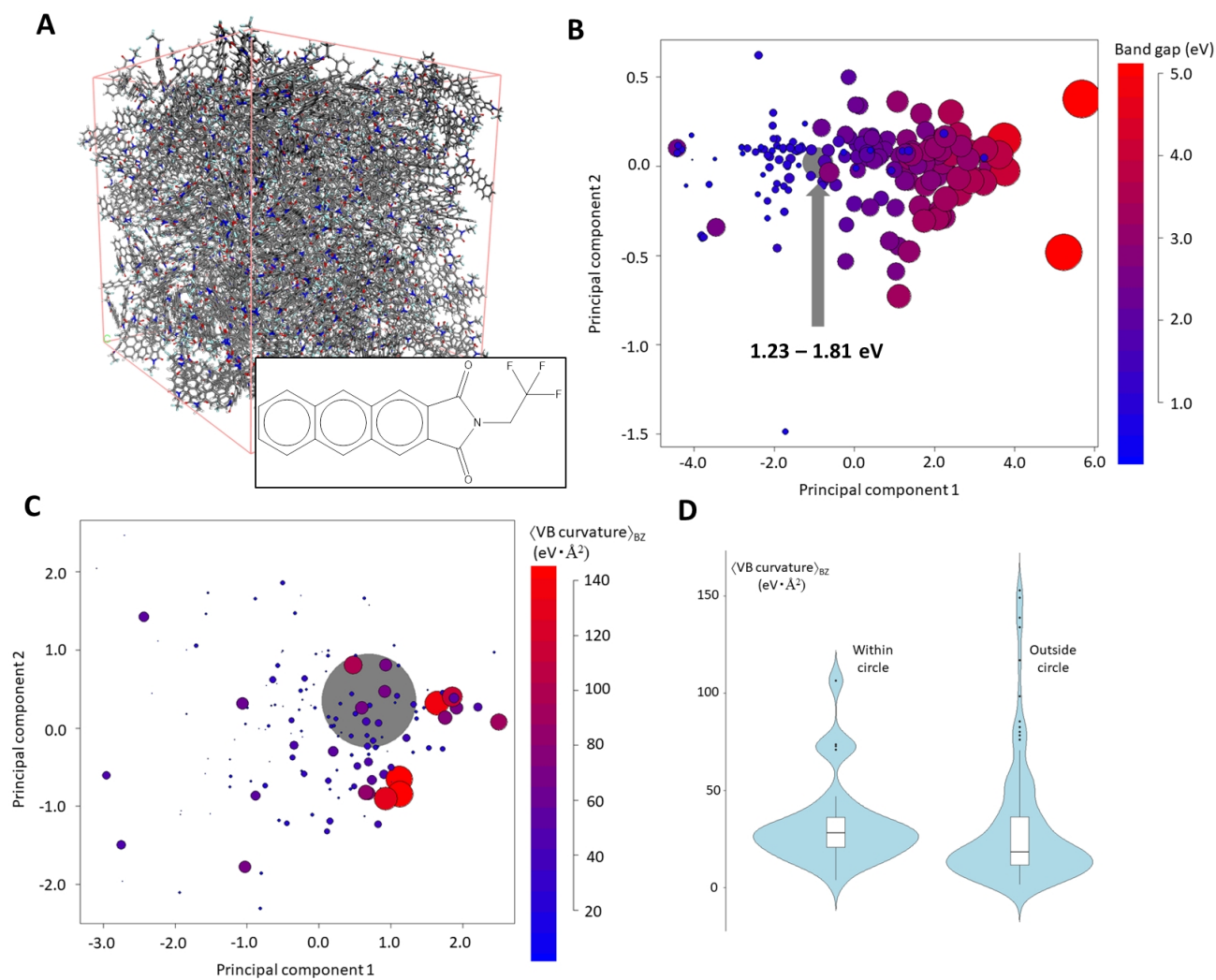

**Figure 4. Evaluation of a new organic semiconducting material.** (A) Molecular structure of molecule **1** (insert) and an atomic model for the amorphous **1** phase. Molecule **1** was obtained from the procedure described in Figure 3. Image produced with the Materials Studio Visualizer software [61]. (B) Estimation of the band bap of molecule **1** in a crystalline phase. Each point in the plot corresponds to one of 179 organic crystals in the database. The principal components are linear combinations of the molecule and dimer energy gaps from the crystal. The colors and radii of the points correspond to their (absolute) band gaps. On the basis of the energy gaps of the molecules and dimers extracted from the amorphous **1** phase, the point for crystalline molecule **1** is expected to lie within the region indicated by the grey circle (indicated by the arrow). (C) Estimation of the Brillouin-zone-averaged valance band curvature (⟨VB curvature⟩$_{BZ}$) for molecule **1** in the crystalline phase. Each point corresponds to one of 152 organic crystals from the database, and colors and radii correspond to ⟨VB curvature⟩$_{BZ}$. The principal components are linear combinations of log-transformed dimer transfer integrals. On the basis of transfer integrals calculated from dimers extracted from the amorphous **1** phase, the point for crystalline molecule **1** is expected to lie in the region indicated by the grey circle. (D) Distribution of ⟨VB curvature⟩$_{BZ}$ for organic crystals inside the grey circle of figure C (left) and outside (right), as illustrated by violin plots (sky-blue) and box-and-whisker plots (white boxes). In the left-hand plot, the interquartile range (white box) represents our estimate of ⟨VB curvature⟩$_{BZ}$ for crystalline molecule **1**. Figures produced using R and the ggplot2 package [51, 63].

Our estimation of the band gap for molecule **1** in the crystalline state is shown in Figure 4B. In this plot, each of the points corresponds to one of 179 organic crystals selected from our database. These crystals were selected at random, under the condition that they contain only one type of molecule. The color and sizes of the points correspond to the (absolute) band gaps of these crystals. The points are positioned according to their projections onto the two principal components. These principal components were constructed from the energy gaps of the highest-occupied and lowest-unoccupied orbitals of the individual molecules and the three lowest-energy dimers extracted from the crystals (see Methods section "Principal component analysis" for details). In this plot, the band gaps generally increase as we move from the left- to the right-hand side. This suggests that these two principal components (particularly the first principal component) are good descriptors for the crystal-phase band gap.

The grey circle (indicated by the arrow) in Figure 4B is our estimate for the location of molecule **1** in the crystalline state. This circle was estimated by repeating the following procedure multiple times: from the set of dimers extracted from amorphous **1**, select three at random, calculate the relevant energy gaps, and place a point on Figure 4B according to the principal components. The idea behind this procedure is that, if iterated enough times, it should identify three dimers which appear in an energetically stable crystal polymorph of molecule **1**. The center of the grey circle indicates the mean of the point locations, and the radius shows the outer limits of the distribution (usually, the distribution would be represented by a circle of radius equal to one standard deviation from the central point, however in this case the distribution is quite tight, making such a circle difficult to discern). The organic crystals in the region of the grey circle have band gaps in the order of 1.23 – 1.81 eV, suggesting that the band gaps of the stable crystal polymorphs of molecule **1** lie within this region as well. This compares well with the (average local) band gaps seen in the organic crystals in the target region in the ESC (ranging between 1 – 2 eV), and suggests that we succeeded to predict a molecule with the desired crystal-phase band gap.

We apply a similar procedure to estimate the average band curvature for molecule **1** in the crystalline state. The result is shown in Figure 4C. In this plot, each point corresponds to one of 152 randomly selected organic crystals. The size and colours of the points correspond to the average band curvature for the crystal. Here, the principal components were obtained by computing the hole transfer integrals for the two most energetically stable dimers in each crystal (see Methods sections "Transfer integral calculations" and "Principal component analysis" for details, and Supporting Information 4 for examples of these dimers). While the correlation between these principal components and average band curvatures is weaker than the one seen above, a general tendency for crystals with higher curvatures to appear on the middle-right section of the plot can be discerned. In order to estimate the location of crystal-phase molecule **1** in this plot, we performed the same procedure as reported above, this time sampling from only the 40 % most energetically stable dimers extracted from amorphous **1**. The distribution of points obtained from this procedure is indicated by the grey circle, whose center and radius are equal to the mean location of the points and the standard deviation of the points from the mean, respectively. Compared to the previous case, the distribution of points is more broad. This broadness likely reflects the weaker correlation between the principal components and the average band curvatures, however it could also be indicative of a highly variable band curvature across the various crystal polymorphs of molecule **1**. In Figure 4D we plot the distribution of the average band curvatures for organic crystals contained within and outside of the grey circle. On the one hand, for the case of crystals outside of the grey circle, the tails of the distribution extends to larger values. Thus, we would not expect for the crystal polymorphs of molecule **1** to display band curvatures comparable to the best organic crystals in our database. On the other hand, within the grey circle, the interquartile range (white box) for the curvature distribution is compressed and shifted upwards towards higher values. The band curvatures for the crystal polymorphs of molecule **1** are therefore expected to be better, on average, then the curvatures exhibited by other crystals in the dataset. This suggests that our design procedure for molecule **1** succeeded to harness molecular structure information from pentacene, ZnPc, $C_{60}$, and other high hole mobility organic crystals contained in the target region.

## 6. Discussion and conclusions

Organic crystals are a challenging target for machine learning (ML)-based design strategies due to the complex relationship that exists between molecule structures and crystal-phase electronic properties. This complex relationship is difficult to express *via* regression models, and limits the effectiveness of standard data-driven design strategies when applied to these materials. In this paper, we presented an alternative approach for designing organic crystals which avoids regressing crystal state electronic properties onto molecule structure-derived variables. At the center of this approach is a so-called electronic structure chart (ESC), a graphical representation of an organic crystal database in which the crystals are positioned according to their bulk electronic properties. Having constructed the ESC, new organic crystals can be designed in three steps. First, one identifies a region of the ESC containing organic crystals with desired electronic properties. Second, one constructs a classifier model which predicts whether a candidate molecule will belong to this region, and hence whether it will express the desired electronic properties, in the crystalline state. Third, one generates a list of candidate molecules and screens them using the classifier model to identify promising cases. The use of a relatively simple classification model in this screening step is highly advantageous for the data-driven design of organic crystals.

While the effectiveness of this ESC-based design strategy should be clear from section 5 above, it has three shortcomings that should be addressed in future work. The first shortcoming is that the design strategy itself is interpolative, meaning that it only predicts materials with electronic properties already contained in the ESC and the dataset. While the importance of this shortcoming should not be dismissed, it is beyond the scope of this work to address here. In fact, all ML-based design strategies suffer this problem, as ML-derived models are not designed to make predictions for cases which substantially differ from the training data. This shortcoming is therefore a general problem which needs to be broadly addressed by all practitioners of material informatics. The second shortcoming concerns the ESC itself. In this paper we have constructed the ESC using electronic fingerprints, which are essentially band structure-derived descriptors. While band structures summarise a wealth of useful information on bulk electronic properties, including the degree of molecular orbital overlap between neighboring molecules, it is common knowledge that charge transport in organic crystals cannot be adequately modeled by band theory. The molecules within an organic crystal are highly dynamic at room temperature, and charge-phonon coupling usually makes substantial contributions to charge transport in these materials. While the local arrangement of organic crystals within our ESC is reflective of the degree of HOMO overlap between neighboring molecules (through the local curvature), it may not be reflective of actual hole mobilities due to the absence of charge-phonon coupling in our electronic fingerprints. However, while this point is important from the point-of-view of designing organic crystals with specific room-temperature charge mobilities, it does not undermine our materials design framework *per se*. If our electronic fingerprints could be generalised to include charge-phonon coupling, then a revised ESC could be constructed *via* the Jansen-Shannon divergence (equation (1)) and the tSNE method as described above. The same design protocol presented in Figure 3 could be applied as well. The question of how to incorporate charge-phonon coupling into the electronic fingerprints without incurring large computational costs should therefore be pursued as future research. The third shortcoming concerns the organisation of the ESC. As was demonstrated in section 3, the ESC is organised on two levels. On a global level, the arrangement of organic crystals in the ESC reflects band gaps, and on a local level the arrangement reflects band curvatures at a nearly constant band gap size. This can be inconvenient for designing new materials. Unless the target region of the ESC is very small and mainly contains organic crystals with similar valance band curvatures, the resulting classifier model will mainly screen molecules on the basis of crystal-phase band gaps rather than band curvatures. In essence, our design strategy puts more 'weight' on band gaps than on valance band curvatures. While this problem might be remedied by shrinking the size of the target region, it will become difficult to construct an accurate classifier model if the target region becomes too small. Future research should therefore investigate why the ESC is organised on two levels, and whether the electronic fingerprints could be improved so that an ESC organised on a single level can be achieved.

The embrace of a data-centric culture by materials scientists is only in its early stages, and efforts to create machine learning-based material design strategies will continue until all major classes of materials are accounted for. Our work has presented a new and effective strategy for the case of organic crystals, the novelty of which lies in the use of an electronic structure chart to simplify the construction of machine-learning models. We hope that design strategies such as ours would transition from a computational curiosity to a routine tool used by synthetic chemists and other materials scientists in the near future.

**Methods**

*Database generation*

The crystal structures for our database (cif files) were obtained from the Crystallography Online Database (COD) [25] in three steps. In the first step, a list of URLs for all C and H-containing crystal structures in the COD was obtained using the COD search feature. In the second step, multiple batches of around 100 structures were selected at random and downloaded. For each batch, a visual inspection was performed to remove structures which contained artifacts such as thermal disorder and under-coordinated atoms. The remaining structures were saved in VASP (POSCAR) format. The second step was repeated multiple times, resulting in a database of 5472 structures. In the third step, 33 additional compounds which are of high practical importance (because they exhibit high hole mobilities or are reasonable targets for synthesis) were selected from the literature and added to the database (see Supporting Information 1 for details). This resulted in a database of 5505 compounds. Note that the additional 33 compounds contained 7 inorganic semiconductors (such as GaAs and InAs) and graphene as well, which were added in the hopes of discovering organic semiconductors with very high hole mobilities; in the end, no such materials were discovered.

For each structure in the database, band structure calculations were performed using DFT as implemented in the Vienna Ab Initio Simulations package (VASP) version 5.4.4 [48]. These calculations used the rev-vdW-DF2 exchange-correlation functional [49], PAW-PBE pseudopotentials [50], 650 eV basis set cut-offs, Gaussian smearing with 0.2 eV smearing widths, and non-self-consistent electron densities computed on 2 x 2 x 2 Γ-centered k-points grids. Brillouin zone paths were generated using AFLOW [37, 38], with 10 k-points between each high-symmetry point. Because relaxation of a single molecular crystal can demand several days of computation, band structure calculations were performed directly on the experimental structures obtained above. These high-throughput band structure calculations were managed using a custom R script [51].

*Electronic fingerprints and electronic structure chart generation.*

To generate the electronic fingerprint for a given band structure, 100 k-points were sampled randomly (with replacement) from the Brillouin zone path. At each k-point $k$ in the sample, the local curvature $c_k = \partial^2 E_{vb}(k)/\partial k^2$ and local energy gap $\varepsilon_k = E_{cb}(k) - E_{vb}(k)$ were computed, where $E_{vb}(k)$ is the energy of the highest occupied ('valance') band and $E_{cb}(k)$ the energy of the next band (the 'conduction' band) at that point. This resulted in a sample of 100 pairs ($\varepsilon_k$, $c_k$). The electronic fingerprints were plotted on 10 x 10 grids, as shown in Figure 1 I – L. The electronic fingerprints did not discernibly change when larger samples of k-points were used.

In order to compute the Jansen-Shannon divergence between a pair of band structures (equation 1), their electronic fingerprints were first plotted on a common grid. 10 x 10 grids were used, and the grids were specified so that they extended from the minimum to the maximum of the local curvatures and energy gaps exhibited by the pair. The Kullback-Leibler divergences in equations (1) and (2) were computed using the discretized formula

$$k(P, M) = \sum_i P(\epsilon_i, c_i) \log_{10} \frac{P(\epsilon_i, c_i)}{M(\epsilon_i, c_i)}$$

where the sum runs over all grid points, which are indexed by $i$, and $P$ and $M$ are the electronic fingerprints for the two band structures under consideration. The use of discrete grids and discretized Kullback-Leibler divergences is preferred to their continuous analogues, because the unavoidable use of discrete k-point paths makes it difficult to obtain a good approximation to a continuous probability distribution. These calculations were performed using a custom R script.

The electronic structure chart in Figure 2A and B were obtained using the Rtsne package for R [52] with a perplexity parameter of 30.

*Autocorrelation functions*

The autocorrelation functions (ACFs) in Figure 2C were computed according to the formula

$$\rho(k) = \frac{\langle (g(X_0) - \mu_0)(g(X_k) - \mu_k) \rangle}{\sigma_0 \sigma_k},$$

where

$$\mu_k = \langle g(X_k) \rangle,$$

and

$$\sigma_k = \left\langle (g(X_k) - \mu_k)^2 \right\rangle^{\frac{1}{2}}.$$

Here, $X$ represents a random walk (Markov chain) which hops between points in the tSNE plot. $X_k$ is the point on which the random walk resides after making $k$ hops. $g(X_k)$ is the value of the property of interest (average local gap, average local curvature, or random noise) for this point. The angular brackets represent the expected value. The term in the numerator is interpreted as the correlation of $g(X_k)$ with $g(X_0)$, the value of the property of interest at the initial position of the random walk.

In this work, this random walk was simulated such that, for a given point in the tSNE plot, a hop was permitted with equal probability to any other point within a cut-off radius of 1.0. The ACFs were computed according to the above equation by simulating 10,000 random walks independently of each other, each initiated from a randomly selected point. Simulations were performed within R using a custom script [51].

*Prediction of candidate molecules*

For each crystal structure in the database, supercells were generated using the ASE library [53]. Molecule structures were converted to SMILES strings using OpenBabel [54]. Default settings were used to construct the regression models with the XGBoost library. iQSPR was implemented as described in the methods sections of references [45] and [46]. All calculations were performed in within Python and managed using a custom Python script.

*Generation of amorphous **1***

800 copies of molecule **1** were placed at random into a 100 x 100 x 100 Å box using the Packmol package [55]. An amorphous phase was then generated in the following steps. In the first step, a geometry optimization was performed. In the second step, velocities were assigned to each atom by sampling at random from a Boltzmann distribution at 600 K, and a sequence of 100 ps-long molecular dynamics simulations were performed at 1000 atm and temperatures of 500, 400, 600, 700, 600, 500, 600, 600, 250, and 300 K, respectively. Another simulation at 1000 atm and 300 K was then performed for 500 ps in order to induce molecule agglomeration. In the third step, a sequence of 2 ns-long simulations were performed at 100 atm at temperatures of 300 K, 250 K, 300 K, and 200 K in order to relax the molecule geometries and local environments around each molecule. In the fourth step, two 2 ns-long simulations were performed at 1 atm and temperatures of 200 K and 200 K, respectively, in order to induce the formation of amorphous **1**. In the final step, the velocity vectors for each molecule were examined to confirm the absence of high-energy molecules. Each molecular dynamics simulation was performed using 2 fs time steps, and the simulation box was allowed to relax in each case. Simulations were performed using the GAFF2 force field [56] and the GROMACS package [57].

*Dimer extraction*

Dimers were extracted by first forming 4 x 4 x 4 supercells from the original crystal structure files and then identifying individual molecules within the supercell.

For a given supercell, the individual molecules were identified using a graph theory strategy. This strategy involves the formation of an adjacency matrix $\mathbf{A} = [a_{ij}]_{n \times n}$, where $n$ is the number of atoms in the supercell. The element $a_{ij} = 1$ if atoms $i$ and $j$ reside within a cut-off distance of each other, and is zero otherwise. Molecules were then identified by tracing out paths within the supercell according to the adjacency relations in $\mathbf{A}$. This resulted in a mixture of full molecules and fragments, where the latter arise from molecules extending across the supercell boundaries. The fragments were ignored in subsequent steps.

Given the molecules extracted from the supercell, dimers were identified as follows. First, for every pair of molecules, a 'dimer distance', defined as

$$d = \frac{1}{n^2} \sum_{i=1}^{n} \sum_{j=1}^{n} |\mathbf{r}_i - \mathbf{r}_j|$$

was calculated. Here, the index $i$ runs over the atoms in one of the molecules in the dimer, and the index $j$ runs over the atoms in the other molecule. $\mathbf{r}_i$ and $\mathbf{r}_j$ denote atom position vectors. The dimer distance is the average of the distances between atoms of different molecules. The six pairs of molecules with the shortest dimer distances were identified and retained for subsequent analysis.

The above steps were performed using custom R and C++ codes. Matrix algebra operations within the C++ code were handled with the Eigen library [58]. For most structures considered here a cut-off distance of 2 Å was satisfactory for the clean extraction of molecules, without severing covalent bonds or including fragments from other molecules. For other cases (particularly structures containing transition metals or heavy main group elements, which contain long bonds), the cut-off distance had to be set specifically.

*Transfer integral calculations*

For a given dimer, transfer integrals were estimated by computing the highest-occupied molecular orbitals (HOMO) of the two molecules in a localised, atom-centered basis. Let

$$\psi_1 = \sum_{i=1}^{r} a_{1i}\phi_{1i},$$

and

$$\psi_2 = \sum_{i=1}^{r} a_{2i}\phi_{2i}$$

represent the respective HOMOs of the molecules in the dimer. Here, we have labeled one of the molecules in the dimer as "1" and the other as "2". $\phi_{1i}$ and $\phi_{2i}$ represent the basis function $i$ on molecule 1 and molecule 2, respectively. $r$ is the number of basis functions per molecule. The transfer integral between $\psi_1$ and $\psi_2$ is defined as

$$T_{21} = \langle \psi_2 | H | \psi_1 \rangle = \sum_{i=1}^{r}\sum_{j=1}^{r} a_{2i}^* a_{1j} H_{ij}$$

where $H$ is the Hamiltonian operator for the dimer and

$$H_{ij} = \langle \phi_{2i} | H | \phi_{1j} \rangle.$$

In this work, the expansion coefficients ($a_{1i}$, $a_{2i}$) and the matrix elements $H_{ij}$ were calculated using DFT. This involves approximating the molecule HOMOs with the highest-occupied Kohn-Sham orbitals. The expansion coefficients are first obtained by performing DFT calculations for the two isolated molecules, and the matrix elements then obtained by performing a calculation for the dimer. These DFT calculations were performed using the FHI-aims package version 171221_1 [59], with the PBE exchange-correlation functional [50], Tkatchenko-Scheffler van der Waals corrections [60], relativistic atomic zora scalar corrections, and "really tight" basis set defaults. Custom R scripts were used to manage the calculations (note that VASP cannot be used to perform these calculations because it is limited to delocalized, plane wave basis sets; consistency between band structures calculated from VASP and FHI-aims was confirmed for several compounds).

Because transfer integral calculations are not a standard feature of the FHI-aims package, we validated our method by comparing our results for pentacene with those reported in reference [61]. See Supporting Information 5 for details.

*Principal component analysis*

The principal components shown in Figure 4B (band gap data) were computed from four variables: the average of the HOMO-LUMO gaps of the two isolated molecules ($g_m$) and the gaps of the three most stable dimers extracted from the structure ($g_1$, $g_2$, and $g_3$, respectively) . This resulted in a data matrix of dimension 179 dimers x 4 variables. The first two principal components were computed to be

$$Q_1 = 0.50 g_m + 0.48 g_1 + 0.51 g_2 + 0.51 g_3$$

and

$$Q_2 = -0.36g_m - 0.10g_1 + 0.84g_2 - 0.39g_3$$

respectively. These principal components accounted for 98.6 % of the variation in the original data matrix.

The principal components shown in Figure 4C (average curvature data) were computed from the two variables $\ln(T_{ij}^1)$ and $\ln(T_{ij}^2)$, where $T_{ij}^1$ and $T_{ij}^2$ are the transfer integrals for the two most stable dimers extracted from the structure. This resulted in a data matrix of dimension 152 dimers x 2 variables. The principal components were computed to be

$$Q_1 = \frac{1}{\sqrt{2}} \left( \ln T_{ij}^1 + \ln T_{ij}^2 \right)$$

and

$$Q_1 = \frac{1}{\sqrt{2}} \left( \ln T_{ij}^2 - \ln T_{ij}^1 \right)$$

respectively, which is the expected result of PCA on two variables. Several other choices of variables were tested, however the resulting principle components did not show a clear correlation with the average curvature values. Since these two principal components were constructed from two variables, they trivially account for 100 % of the variation in the original data matrix.

Principal component analysis was performed in R [51] using the prcomp function.

## Acknowledgments

DP was supported by donations provided by the Daicel Corporation and JSPS KAKENHI grants 21K05003, 19H04574, and 18K14126.

## Conflicts of interest

The authors report no conflicts of interests.

## Author contributions

DP, YK, DI, and MO conceived the study together. DP collected the data, performed the DFT calculations (section 2), and created the electronic structure chart (section 3). YK implemented the materials design method (section 4). DI and MO selected molecule **1** from the output of the materials design method by considering synthetic potential. DP and YK evaluated the crystal state properties of molecule **1** (section 5). DP drafted the paper, and DP, YK, DI, and MO finalized the paper together.

Daniel M. Packwood[1], Yu Kaneko[2], Daiji Ikeda[2], Mitsuru Ohno[2]

[1] Institute for Integrated Cell-Material Sciences (iCeMS), Kyoto University, Kyoto 606-8501, Japan

[2] Daicel Corporate Research Center, Innovation Park (iPark), Daicel Coorporation, Himeiji, Japan

# 1. Crystal structures in the database

1a. 5472 crystal structures from the Crystallography Open Database (COD), listed as COD IDs. See the file "data_cod.txt". Original CIF files can be accessed at http://www.crystallography.net/cod

1b. 33 crystal structures with high electron or high mobilities

| ID in our database | Common name | Reference |
|---|---|---|
| 1544287-1544289 | ChDT | *Adv. Sci.* 5, 2018, 17003717 |
| 1544287-1544289_2 | $C_{10}$ThChDT | *Adv. Sci.* 5, 2018, 17003717 |
| 886144-886147 | DNT-V, | *Adv. Mater.* 25, 2013, 6392 |
| 886144-886147_2 | $C_{10}$-DNT-VW-oDCB | *Adv. Mater.* 25, 2013, 6392 |
| 886144-886147_3 | $C_{10}$-DNT-VW,-TCE | *Adv. Mater.* 25, 2013, 6392 |
| 886144-886147_4 | $C_{10}$-DNT-VV | *Adv. Mater.* 25, 2013, 6392 |
| 9999997 | Graphene | [Created in Materials Studio 2017] |
| 9999998 | Pentacene | *J. Am. Chem. Soc.* **129**, 2007, 10316 |
| 9999999 | $C_{60}$ | *Nature* **353**, 1991, 147 |
| am3026163_si_001 | 2,9-$C_{10}$-DNTT | *ACS Appl. Mater. Interfaces.* **5**, 2013, 2331 |
| am3026163_si_002 | 2,10-$C_{10}$-DNTT | *ACS Appl. Mater. Interfaces.* **5**, 2013, 2331 |
| c3cc47577h2 | $C_{10}$-DNF-VV | *Chem. Commun.* **50**, 2014, 5342 |
| c3cc47577h3 | $C_{10}$-DNT-VW | *Chem. Commun.* **50**, 2014, 5342 |
| c6ra00922k2 | C10-DNBDF-NW | *RSC Adv.* **6**, 2016, 28966 |
| c6ra00922k3 | DNBDF | *RSC Adv.* **6**, 2016, 28966 |
| c6ra00922k4 | C10-DNBDF-NV | *RSC Adv.* **6**, 2016, 28966 |
| c6tc04721a2 | $C_{10}$-TBBT-V | *J. Mater. Chem. C.* **5**, 2017, 1903 |
| cm303376g_si_002 | DNF-W | *Chem. Mater.* **25**, 2013, 3952 |
| cm303376g_si_003 | DNS-W | *Chem. Mater.* **25**, 2013, 3952 |
| cm303376g_si_004 | DNT-W | *Chem. Mater.* **25**, 2013, 3952 |
| Inorg1 | GaAs | materialsproject.org |
| Inorg2 | GaSb | materialsproject.org |
| Inorg3 | Ge | materialsproject.org |
| Inorg4 | InAs | materialsproject.org |
| Inorg5 | InP | materialsproject.org |
| Inorg6 | InSb | materialsproject.org |
| Inorg7 | Si | materialsproject.org |
| ja110973m_si_003 | DPh-NDT4 | *J. Am. Chem. Soc.* **133**, 2011, 5024 |
| ja110973m_si_004 | DPh-BTBT | *J. Am. Chem. Soc.* **133**, 2011, 5024 |
| ja202377m_si_001 | DATT | *J. Am. Chem. Soc.* **133**, 2011, 8732 |
| ja406257u_si_002 | BTAT | *J. Am. Chem. Soc.* **135**, 2013, 13900 |
| ja406257u_si_003 | BBTNDT | *J. Am. Chem. Soc.* **135**, 2013, 13900 |
| ZnPc | α-Zinc phthalocyanine | |

## 2. Example compounds from the electronic structure chart

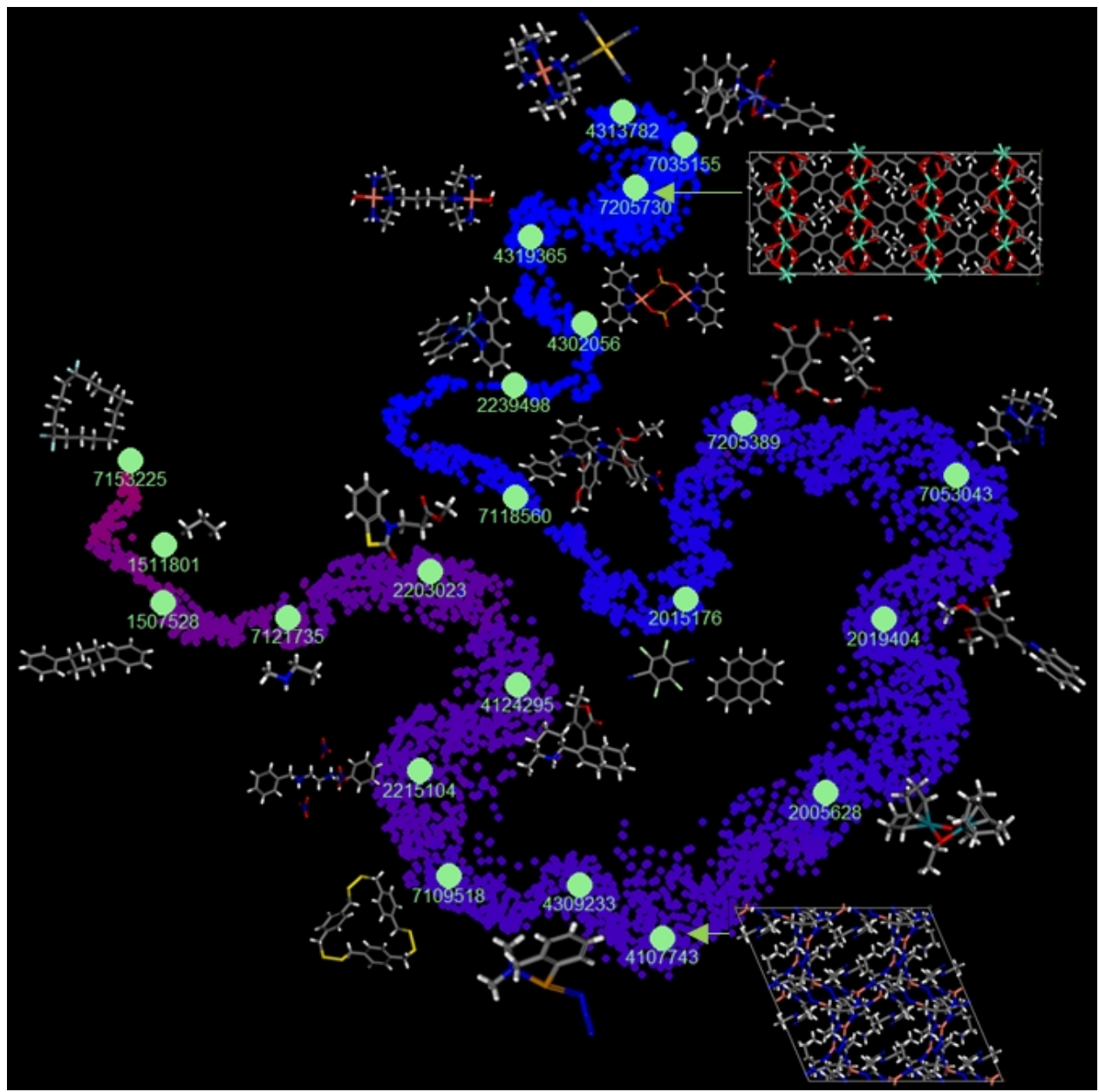

[Above] As for Figure 2A, but with some example compounds indicated. The points are labeled according to their numbers from the Crystallography Online Database. With the exception of 7205730 and 4107743, isolated molecules rather than entire crystal structures are shown for clarity. White = hydrogen, grey = carbon, blue = nitrogen, red = oxygen, yellow = sulfur (in 2203023 and 7109518), green = chlorine (in 2015176), light orange = copper (in 4313782, 4302056, and 4107743), light blue = cobalt (in 7035155), light yellow = gold (in 4313782), orange brown = tellurium (in 4309233), dark yellow = selenium (in 4302056), light blue = nickel (in 2239498), cyan = rhodium (in 2005628), light green = europium (in 7205730), blue-grey = zinc (in 7205389)

## 3. Molecules output from iQSPR

The 248 molecules output from 24 rounds of iQSPR are listed in [SI3.xlsx](SI3.xlsx). Molecule structures, SMILES strings, predicted labels, and measured melting and boiling points are listed.

## 4. Transfer integrals for dimers extracted from amorphous 1

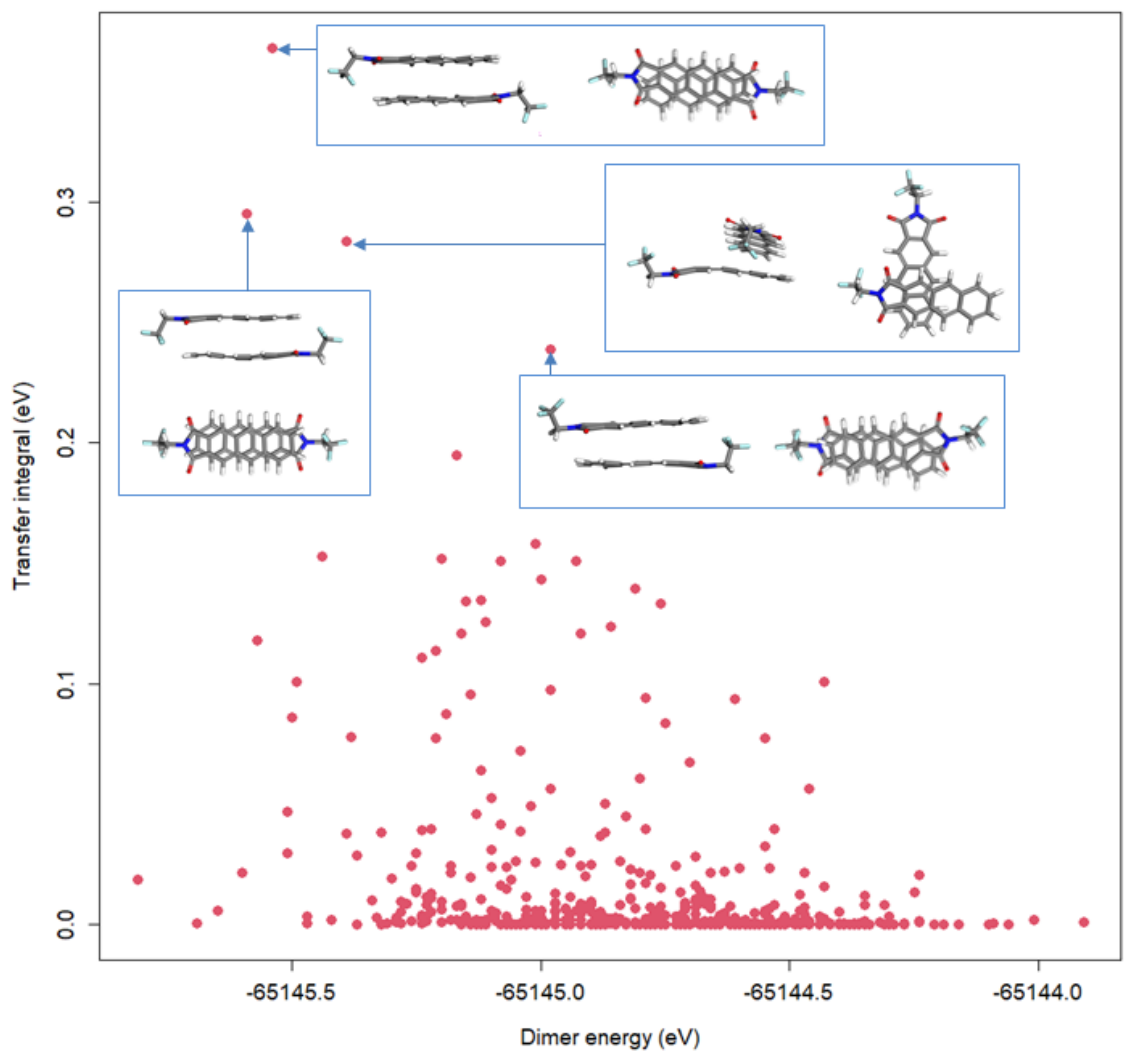

[Above] Energies and (absolute) transfer integrals computed from the 465 dimers extracted from amorphous **1**. Side- and top-views of the four dimers with highest transfer integrals are shown. White = hydrogen, grey = carbon, red = oxygen, blue = nitrogen, aqua blue = fluorine.

## 5. Validation of transfer integral calculations

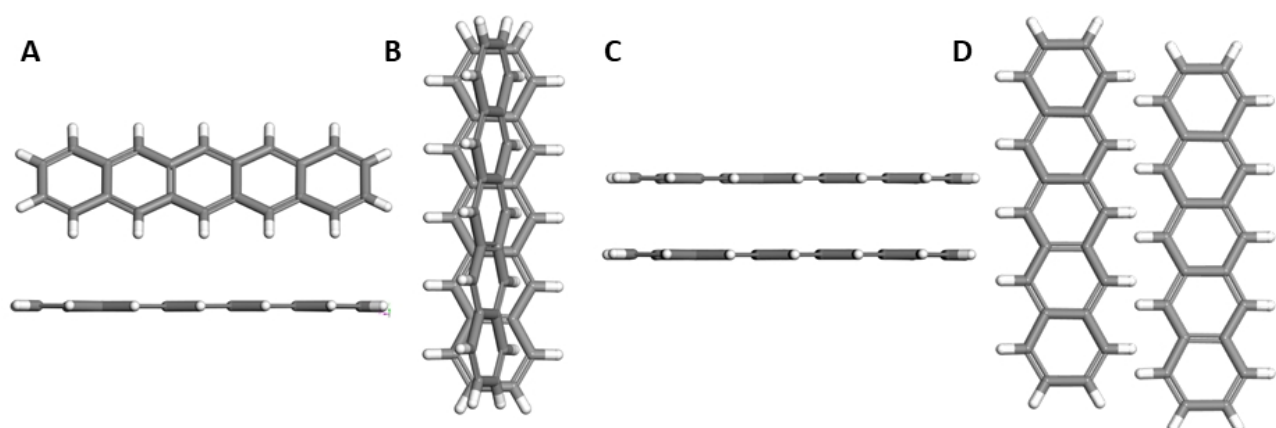

Two dimers extracted from the pentacene crystal. A and B are respectively side and top views of one of these dimers. C and D are views of the other dimer. Our method yields 0.11 eV and 0.07 eV for the transfer integrals of these dimers. These compare to 0.10 eV and 0.06 eV computed by other authors (Nguyen, T. P., Shim, J. H., Lee, J. Y. Density Functional Theory Studies of Hole Mobility in Picene and Pentacene Crystals. *J. Phys. Chem. C.* **119**, 2015, 11301). White = hydrogen, grey = carbon.